\RequirePackage[2020-02-02]{latexrelease}
\documentclass[twocolumn,showpacs,preprintnumbers]{revtex4}
\usepackage{amsmath}
\usepackage{epsfig}
\usepackage{graphicx}
\usepackage{xcolor}
\begin{document}

\title{ Universal behavior of the condensation energy of Superconducting BCS Bose gases}

\author{J. J. Valencia$^1$, M.A. Sol\'is$^2$, Patricia Salas$^3$, I. Ch\'avez$^2$ \\
	$^1$ Universidad Aut\'onoma de la Ciudad de M\'exico, Plantel San Lorenzo Tezonco, \\
	Apdo. postal 09790, 09780 Ciudad de México, Mexico\\
	$^2$ Instituto de F\'isica, Universidad Nacional Aut\'onoma de M\'exico, \\
	Apdo. postal 20-364, 01000 Ciudad de México, Mexico \\
    $^3$ Facultad de Ciencias, UNAM, Apdo. postal 70-542, 04510 Ciudad de México, Mexico} 
\date{Modified: \today / Compiled: \today}

\begin{abstract}

  Using the Boson-Fermion formalism of superconductivity we calculate the   
 condensation energy for several superconductors ranging from conventional to unconventional, or high temperature superconductors (HTSC).  
 It is calculated as the difference between the Helmholtz free energies of the superconducting and the normal state, which  
 is a gas of $N$ attractive electron gas, while the superconducting state is formed by the condensed Cooper pairs taken as composite bosons, coming from
 a fraction 
 of the electrons inside the Debye shell, plus those electrons inside  and  outside the Debye shell that are unable to pair. 
 The energy-momentum relation (dispersion relation) of the composite bosons includes a
 Bardeen-Cooper-Schriffer (BCS) temperature dependent jump at the Cooper pair center of mass $K=0$, $\Delta(T)$, between the ground state and the first excited state, which becomes zero at a temperature $T_B$ (not necessarily the superconducting critical temperature $T_{cs}$).
 After giving the analytic expressions for the internal energy $U(T)$ and the entropy $S(T)$ we obtain the Helmholtz free energy $F(T) = U -TS$ for both the superconducting 
 ($F_s$)
 and 
 the normal states ($F_n$) as functions of temperature, which are used to calculate the condensation energy $E_{cond} = F_s(T) – F_n(T)$. 
 In the search for universalities, we calculate the ratio of the condensation energy at $T=0$ to the Sommerfeld constant  $\gamma_0$
 (the normal state electronic specific heat over the temperature when $T \rightarrow 0$) using two different methods: the Boson-Fermion formalism developed here, as well as an analytical expression deduced from a combination of the BCS and Ginzburg-Landau theories. 
 We find  for the Boson-Fermion formalism $E_{cond}/\gamma_0= 0.252\,T_{c}^{1.997}$, which is the same behavior  described by the experimental fit of  Kim {\it et al.} $E_{cond}/\gamma_0= 0.2\,T_{c}^{2.06}$  \cite{Kim-Tam} and  by the one recently  reported by Tallon {\it et al.} \cite{Tallon2026} for overdoped cuprate superconductors; while for the Ginzburg-Landau-BCS we get the expression $E_{cond}/\gamma_0= 0.236\,T_{c}^{2}$, also in very good agreement with the partifirst method.

\end{abstract}
\maketitle

\section{Introduction}

Since the discovery 
of other high temperature superconductors in addition to cuprates \cite{Bednorz,Kamihara,Budko2015}, the search for universal correlations among their superconducting properties that can be applied to the conventional elements and to the new classes of superconductors, has been in the focus of experimental and theoretical efforts \cite{Kim-Tam,Tallon2026}.

One of the first attempts to find correlations is the famous Uemura's relation, which associates the superconducting critical temperature $T_{cs}$ with the carrier density $n_s$ and the inverse of the penetration length $\lambda$ \cite{Uemura1989}, which led to the Uemura's plots that relate the transition temperature with the Fermi temperature, a diagram that has been updated over the years as new unconventional superconductors appear \cite{Uemura2009}. But, although most superconductors obey a certain pattern in those relations, some of them are obviously out of range, especially the conventional ones, so a universal correlation cannot be inferred from these diagrams. A later attempt follows from the work by Bud'ko {\it et al.}, where a correlation between  $\Delta C_p$  and $T_{cs}^3$ is found for a large variety of iron based superconductors \cite{Budko2009}; however,  conventional superconductors don't fit in this relation, since they obey the BCS universal relation $\Delta C_p = 1.43\, \gamma_0 \, T_{cs}$.

Another attempt to find universal relations is to associate quantitatively the superconducting transition temperature $T_{cs}$ to the Debye temperature $\Theta_D$ (proportional to the energy of the phonons) as in Ref. \cite{Esterlis}, where it is found that  $T_c \leq0.1 \Theta_D$ for most conventional superconductors, where the electron-phonon coupling $\lambda_{e-ph}$ prevails, which are known as adiabatic since they fulfill the condition $\Theta_D/T_F \ll 1$. This relationship has been recently confirmed by applying machine learning models to conventional superconductors, including high-pressure hydrides \cite{Smith}.  
But in order to include unconventional  superconductors (nonadiabatic, with  $\Theta_D/T_F \gg 1$), these relations are  empirically extended in Ref. \cite{TalantsevNano}, where the quotient $\Theta_D/T_F$ is graphed either {\it vs.} $T_{cs}$ or $\lambda_{e-ph}$ for several families of superconductors, finding that most of them lie in a moderately strong nonadiabatic regime ($0.025 \le \Theta_D/T_F \le 0.4$).



In this work we propose two different analytic ways to find a universal relation between the ratio of the condensation energy at $T=0$ over the Sommerfeld constant, by using the Boson-Fermion formalism of superconductivity and from the conjunction of the BCS and the Ginzburg-Landau (GL) expressions.

Experimentally, the condensation energy is usually obtained from the isobaric specific heat measured for the normal and the superconducting states, which is analytically integrated to get the corresponding entropies, which in turn are used to get the condensation energy  \cite{Kim-Tam,Xing2014}. Another way to get the condensation energy is by using the measured thermodynamic critical field $H_{c0}$, where available. 
In their work, Kim {\it et al.}  use the integral of the difference between the electronic  normal and  superconducting experimental entropies, $\int_{0}^{T_c} (S_n - S_s)$, to obtain the condensation energy  for several unconventional superconductors; while they use the critical field  $H_{c0}^2/8\pi$ for conventional ones \cite{Kim-Tam}. Additionally, they show that the ratio of the obtained  values for the condensation energy over the experimental $\gamma_0$  is  $0.2 \,T_{c}^{2.06}$ for the wide range of superconductors discovered at the time their analysis was done. 
Kim's work cites Loram's {\it et al.} \cite{Loram2001} earlier work on cuprate superconductors, where it was experimentally observed that for some overdoped cuprates the relation between the condensation energy over the normal state specific heat over the temperature ($\gamma_n$) is proportional to $T_{c}^2$. Recently,  Tallon {\it et al.} \cite{Tallon2026} showed that this behavior is valid for most overdoped cuprates; however, in the mentioned works, Refs.  \cite{Loram2001,Tallon2026},  $\gamma_n$ is used instead of $\gamma_0$, so the constant of proportionality  is different for each cuprate.
Also, the expression mentioned above used to obtain the condensation energy is susceptible to improvement, since it implies that the difference between the superconducting and normal chemical potentials is negligible, which has been shown, that in some cases, it is of the same order of magnitude as the condensation energy  \cite{ChavezPhysicaA}.


A theoretical attempt to relate the condensation energy at $T=0$ over $\gamma_0$ with ${T^2_{c}}$ has been developed in \cite{Shaginyan}, where they extend the  correlation  to conventional and unconventional superconductors based on the theory of fermion condensation  at $T=0$, generating a superconducting state with flat bands for both kinds of superconductors, which leads to a finite effective mass for unconventional superconductors, making them BCS-like. However, they are not accurate in the constants involved in their development.

In the Boson-Fermion formalism  \cite{flee,GBEC}  superconductivity is taken as a Bose-Einstein condensation (BEC) of Cooper pairs treated as composite bosons formed from a fraction $f\,N_{\omega}$,
with $f<1$ \cite{Salas} and $N_{\omega}$ the electrons inside the Debye shell, plus those electrons inside  and  outside the Debye shell that are unable to pair, while the normal state is formed by a gas of $N$ interacting electrons.
In this scenario, the condensation energy is calculated as the difference between the Helmholtz free energy (HFE) 
$F(T) = U -TS$ of the superconducting 
($F_s$)
and 
the normal states ($F_n$) as functions of temperature, which are used to calculate the condensation energy $E_{cond} = F_s(T) – F_n(T)$. 

A three-dimensional (3D) Bose gas (BG) with a generalized dispersion relation plus a constant jump $\Delta_0$, presents a Bose-Einstein critical temperature $T_c$, which is jump magnitude, dimensionality $d$ and energy-momentum exponent $s$ dependent, as shown in \cite{Martinez-Herrera}. When the jump magnitude is zero, the BG case with a generalized dispersion relation is  recovered, whose thermodynamic properties are reported in Ref. \cite{Aguilera-Navarro}. The
3D ideal Bose gas (IBG) with $d=3$ and $s=2$ is a particular case, whose thermodynamic properties, as well as its Bose-Einstein critical temperature $T_0$, are straightforwardly  recovered \cite{Pathria}.


In this work we analyze the influence of an energy-momentum relation (dispersion relation) of the composite bosons, which includes a
Bardeen-Cooper-Schriffer (BCS) temperature dependent jump at the Cooper pair center of mass $K=0$, $\Delta(T)$, between the ground state $\varepsilon_0$ and the first excited state, which becomes zero at temperature $T_B$ (not necessarily the critical temperature) plus a quadratic term $\varepsilon _{0}+\Delta(T) +\hbar^{2}K^{2}/2m_b$, leading to two special cases: without jump, $\Delta(T)=0$ if $T_B =0$, i.e. the ideal Bose gas, and with a  constant jump, $\Delta(T)=\Delta_0$ if $T_B$ goes to infinity. 
In Sec. II. we give the number equation for the boson gas, from which the condensation critical temperature and the chemical potential may be obtained,  the boson internal energy, the boson entropy, and the Helmholtz free energy for the boson and fermion gas. 
In Sec. III  the condensation energy $E_{cond}$ is given using the Boson-Fermion formalism, and separately we obtain an analytic expression for the condensation energy $E_{GL-BCS}$ deduced from the Ginzburg-Landau \cite{GL} merged with the BCS theories. In Sec. IV we calculate the ratio  $E_{cond}/\gamma_0$ for each one of the expressions previously obtained, 
showing that they both present a universal behavior as functions of the superconducting critical temperature $T_{cs}$ 
for many superconductors ranging from conventional to unconventional, obtaining the relation $E_{cond}/\gamma_0=0.252\, T_{c}^{1.997}$ 
with the Boson-Fermion formalism and $E_{GL-BCS}/\gamma_0=0.236\, T_{c}^{2}$ with the Ginzburg-Landau-BCS expression deduced here, where we have taken $T_{cs} = T_{c}$. 
In Sec. V we present our conclusions.


\section{Properties of a Cooper-pairs Bose gas}


Our system is a 3D infinite boson-fermion mixture coming from an original Fermi gas of $N_e$ interacting electrons of mass $m_e$ of  which only a fraction is inside the Debye shell
are paired to form a Cooper pair gas of $N_b$ composite bosons with mass $m_b = 2\, m_e$. 

The boson energy-dispersion relation, as a function of the magnitude of their center of mass momentum $\hbar K$, is given by \cite{annals}    
\begin{equation}
	\varepsilon _{_K}=\left\{
	\begin{array}{c}
		\varepsilon _{0}{\ \ \ \ \ \ \ \ \quad \ \quad \ \qquad \ \ \ \ \ \ \ \ \ \mbox{if}\ \ \  }K=0 \\
		\varepsilon _{0}+\Delta(T) +\hbar^{2}K^{2}/2m_b{\ \ \ \mbox{if} \ \ \ }K>0%
	\end{array}%
	\right. ,  
	\label{rdg}
\end{equation}
where $\Delta (T)$ is the temperature dependent energy gap between the ground state energy $\varepsilon_0$ and the first excited state  $\varepsilon_0+\Delta(T)$. 
Although the jump $\Delta (T)$ may have different forms, in order to compare with BCS theory results, here we use a temperature dependent BCS-type gap, denoting as $T_B$ the temperature where the gap becomes zero,  
%
which in general may be different from the BEC critical temperature $T_c$,  
as
\begin{equation}
	\Delta (T) =\left\{
	\begin{array}{c}
		\Delta_0(1 - T/T_B)^{1/2}\ \ \ \    {   \mbox{if}\ \ \  }T < T_{B} \\
		\quad 0 \quad \quad \quad \quad \quad {\    \, \,\, \quad  \   \ \mbox{if} \ \  } \  T \geq T_{B},%
	\end{array}%
	\right.   
	\label{gapTB}
\end{equation}
where $\Delta_0$ is the magnitude of the gap at $T = 0$. 
We notice two limiting cases: when $T_B = 0$ we recover the case where  $\Delta(T) = 0$ of an ungapped ideal Bose gas \cite{Pathria}, while when $T_B \rightarrow \infty$ we recover the case of the ideal Bose gas plus a constant gap $\Delta(T) = \Delta_0$ \cite{Martinez-Herrera}.
From here on we will be using the definitions $\Delta(T_c) \equiv \Delta_c$ and $\Delta(T) \equiv \Delta$.

The electrons that make up Cooper pairs come from an ideal Fermi gas whose dispersion relation is quadratic, $\varepsilon_k =\hbar^2 k^2/2m_e$.

\subsection{Bose-Einstein condensation critical temperature}


The boson number $N_b$ for a given finite temperature is distributed between the energy ground and the excited states, i.e., $N_b=N_{b0}(T)+ N_{be}(T)$, with 
\begin{equation}
	N_{b0}(T)={\frac{1}{e^{\beta \lbrack \mathbf{\varepsilon }_{0}-\mu (T)]}-1}}
	\label{N0}
\end{equation}
the particles in the energy ground state and
\begin{equation}
	N_{be} = \sum_{\mathbf{K\neq 0}}n_{_{K}}  = \sum_{\mathbf{K\neq 0}}{\frac{1}{e^{\beta \lbrack \mathbf{\varepsilon }%
				_{_{K}}-\mu (T)]}-1}} \   
	\label{Ne1}
\end{equation}
the particles in the excited states,    
where $\beta \equiv 1/k_{B}T$, and $\mu (T)\leq \varepsilon _{0}$ the bosonic chemical potential, which is $\varepsilon _{0}$ for $T \leq T_{c}$. 

The BEC critical temperature $T_c$ is the smaller temperature for which  
$N_{b0}(T_{c})/N_b \simeq 0$, $N_{be}/N_b\simeq 1$ and $\mu(T_c)=\varepsilon_{0}$, 
and after some algebra we arrive to \cite{annals}
%
\begin{eqnarray}
	N_{be} &=&\frac{L^{3}(2m_b/\hbar ^{2})^{3/2}}{4\pi ^{2}\beta^{3/2}}%
	\int_{0}^{\infty }{\frac{\xi ^{1/2}d\xi }{e^{-\beta (\mu(T)-\varepsilon_0-\Delta)} \ e^{\xi }-1}}  \nonumber \\
	&{=}&\frac{L^{3}(2m_b/\hbar ^{2})^{3/2}}{4\pi ^{2}\beta^{3/2}}%
	g_{3/2}(z_{1})\Gamma (3/2).
	\label{Ne2}
\end{eqnarray}
where $g_\nu (z)$ is the Bose function of order $\nu$ and $z_{1} \equiv \exp[\beta(\mu(T) - \varepsilon_0 - \Delta)]\leq 1.$ 
Therefore, for $T= T_c$ we obtain

\begin{equation}
	N_b = \frac{L^{3}(2m_b/\hbar^{2})^{3/2}}{4\pi ^{2}\beta _{c}^{3/2}}%
	g_{3/2}(z_{1c})\Gamma (3/2)
	\label{Tc}
\end{equation}
with $z_{1c} \equiv \exp[-\beta _{c}\Delta_c]\leq 1.$ 
For $\Delta = 0 $, $z_{1c} = 1$ so Eq. (\ref{Tc}) 
becomes 
\begin{equation}
	N_b = \frac{L^{3}(2m_b/\hbar^{2})^{3/2}}{4\pi ^{2}\beta _{0}^{3/2}}%
	\zeta (3/2) \Gamma (3/2)
	\label{T0}
\end{equation} 
with $\beta _{0}=1/k_BT_0$ and $T_0$ the BEC critical temperature for an ungapped ideal Bose gas (IBG)
with a number density equal to that of our system.

From the ratio between (\ref{Tc}) and (\ref{T0}), we obtain the expression between 
$T_c$ and 
$T_0$ 
\cite{annals} 
\begin{eqnarray}
	1 &=&\frac{\beta _{0}^{3/2}g_{3/2}(z_{1c})}{\beta _{c}^{3/2}\zeta (3/2)}%
	\qquad \nonumber\\
	{\mbox{so} \qquad }T_{c} &=&\left( \frac{\zeta (3/2)}{g_{3/2}(z_{1c})}\right)
	^{2/3}T_{0},
	\label{eq:Tc/T0}
\end{eqnarray}
where $\beta_c = 1/k_B\,T_c$,
which in terms of $T_F$ becomes
\begin{equation}
	T_{c} =\left( \frac{\zeta (3/2)}{g_{3/2}(z_{1c})}\right)
	^{2/3}0.218033\,T_F. \label{eq:Tc/TF}
\end{equation}


Since we are deriving the condensation energy, we first obtain the expressions for the internal energy $U(T)$, the entropy $S(T)$, and the corresponding HFE $F=U-TS$, for both the normal and the condensed states. 


\subsection{Boson internal energy}

For this system, the bosonic internal energy $U_b(T)$ divided by $T$ as a function of the temperature is given by (see Eq. (21) of \cite{annals})  
\begin{equation}
	\begin{split}
		&	\frac{U_b(L^{d},T)}{N_bk_B T}=\frac{\mathbf{\varepsilon }_{0}}{k_B T} \\
		& + \left[\frac{\Delta }{k_B T}\frac{g_{3/2}(z_1)}{g_{3/2}(z_{1c})}  
		+\frac{3}{2}\frac{g_{5/2}(z_1)}{g_{3/2}(z_{1c})}\right]\left(\frac{T}{T_c}\right)^{3/2}.
		\label{UkBT}
	\end{split}
\end{equation} 
For $T>T_c$ the number of excited particles is equal to the total number of particles, so dividing Eq. (\ref{Ne2})
by Eq. (\ref{Tc}) we have \cite{annals}
\begin{equation}
	T^{3/2}g_{3/2}(z_1)=T_c^{3/2}g_{3/2}(z_{1c}),
	\label{zg3/2}
\end{equation}
which we introduce in the internal energy Eq. (\ref{UkBT}) to obtain
\begin{equation}
	\begin{split}
		\frac{U_b(L^{d},T)}{Nk_B T}=\frac{\mathbf{\varepsilon }_{0}}{k_B T}
		+ \frac{\Delta}{k_B T}  
		+\frac{3}{2}\frac{g_{5/2}(z_1)}{g_{3/2}(z_1)}.
		\label{UkBT2}
	\end{split}
\end{equation}

\subsection{Boson entropy}

 
The entropy is frequently obtained from 
the specific heat at constant volume as 
\begin{equation}
	\frac{S_b}{N_bk_B}=\int_0^T\frac{C_V}{N_bk_BT}dT
	\label{sCv1}
\end{equation}
 (see \cite{Pathria}, Eq. (27) of Appendix H). However, by substituting in the last equation the isochoric specific heat as the derivative of the internal energy $U(T)$ with respect to the temperature and integrating by parts, we obtain
\begin{equation}
	\frac{ S_b}{N_bk_B}=\left. \frac{U_b}{N_bk_BT}\right|_{0}^{T}+\int_0^T \frac{U_b}{N_bk_BT^2}dT.
	\label{su}
\end{equation}



Substituting (\ref{UkBT}) in (\ref{su}) for $T<T_c$, and after some algebra, we obtain 
\begin{equation}
	\begin{split}
		\frac{S_b}{N_bk_B}T_c^{3/2}g_{3/2}(e^{-\beta_c\Delta_c})=&
		\left[ \frac{\Delta }{k_B T} g_{3/2}(e^{-\beta\Delta}) \right. +\\ \left. \frac{3}{2}g_{5/2} (e^{-\beta\Delta}) \right] T^{3/2}
		+&\int_0^T \frac{\Delta }{k_BT^{1/2}}g_{3/2}(e^{-\beta\Delta})dT \\
		 + \frac{3}{2}\int_0^T T^{1/2}g_{5/2}(e^{-\beta\Delta}) dT  ,
	\end{split}
	\label{s15}
\end{equation}
where we have used that $\mu=\varepsilon_{0}$.

Doing the last integral on the right hand side of (\ref{s15}) by parts, we obtain
\begin{equation}
	\begin{split}
		\int_{0}^{T}T^{1/2}g_{5/2}(e^{-\beta\Delta})dT =  \left. \frac{2}{3}T^{3/2}g_{5/2}(e^{-\beta\Delta}) \right|_{0}^{T} & \\
		+ \frac{2}{3k_B}\int_{0}^{T} \left[ \frac{d\Delta}{dT} - \frac{\Delta}{T} \right] T^{1/2}g_{3/2}(e^{-\beta\Delta}) dT&,	\end{split}
	\label{IT1/2g5/2}
\end{equation}
\noindent
so Eq. (\ref{s15}) can be rewritten as 
\begin{equation}
	\begin{split}
		\frac{S_b(T)}{N_bk_B}=& \frac{\left(T/T_c\right)^{3/2}}{g_{3/2}(e^{-\beta_c\Delta_c})}\left[\frac{5}{2}g_{5/2}(e^{-\beta\Delta})+\frac{\Delta}{k_BT}g_{3/2}(e^{-\beta\Delta}) \right. \\ 
		&+  \left. \frac{1}{k_BT^{3/2}}\int_0^T T^{1/2}\frac{d\Delta}{ dT}g_{3/2}(e^{-\beta\Delta})dT \right].
	\end{split}
	\label{Smntmtc}
\end{equation}

Here we analyze two special cases:
When there is no gap,  i. e. $\Delta = 0$, we have
\begin{eqnarray}
	\frac{S_b(T)}{N_bk_B} = 
	 (T/T_c)^{3/2} \frac{5 \zeta(5/2)}{2 \zeta(3/2)}, 
	\label{D0}
\end{eqnarray}
\noindent
while for a constant gap $\Delta = \Delta_0$= $k_B T_0$, which is equivalent to taking $T_B \rightarrow \infty$, we get
\begin{eqnarray}
	\frac{S_b(T)}{N_b k_B}= & \frac{\left(T/T_c\right)^{3/2}}{g_{3/2}(e^{-\beta_c\Delta_0})}\left[\frac{5}{2}g_{5/2}(e^{-\beta\Delta_0}) + \frac{\Delta_0}{k_BT}g_{3/2}(e^{-\beta\Delta_0}) \right]. \nonumber \\
\end{eqnarray}

We need to find an expression for the entropy when $T>T_c$, so substituting Eq. (\ref{UkBT2}) in (\ref{su}), and after some algebra we arrive to
\begin{equation}
	\begin{split}
		\frac{\Delta S_b}{N_bk_B}= \frac{S_b(T) - S_b(T_c) }{N_bk_B} = \left. \frac{3}{2}\frac{g_{5/2}(z_1)}{g_{3/2}(z_1)} \right|_{T_c}^{T} \\ +\int_{T_c}^T \frac{1}{k_B T}\frac{d\Delta}{dT}dT 
		+\frac{3}{2} \int_{T_c}^T \frac{1}{T}\frac{g_{5/2}(z_1)}{g_{3/2}(z_1)}dT.
		\label{su2}
	\end{split}
\end{equation}

%
Using relation (\ref{zg3/2}) on the last term on the right hand side of Eq. (\ref{su2})  
and after some algebra, 
we have
\begin{equation}
	\begin{split}
		\int_{T_c}^T \frac{1}{T}\frac{g_{5/2}(z_{1})}{g_{3/2}(z_{1})}dT = \frac{2}{3}\frac{g_{5/2}(z_{1})}{g_{3/2}(z_{1})} - \frac{2}{3}\frac{g_{5/2}(z_{1c})}{g_{3/2}(z_{1c})} \\
		- \frac{2}{3}\frac{\mu - \varepsilon_0 - \Delta}{k_BT} - \frac{2}{3}\frac{ \Delta_c}{k_BT_c}.
		\label{Ig5/2eTg3/2}
	\end{split}
\end{equation}

Substituting this  result in (\ref{su2}) we arrive at the expression for the entropy for the temperature range $T>T_c$ 
\begin{equation}
\begin{split}
		&\frac{S_b(T)-S_b(T^-_c)}{N_bk_B}= \frac{5}{2}\frac{g_{5/2}(z_1)}{g_{3/2}(z_1)} -\frac{5}{2}\frac{g_{5/2}(z_{1c})}{g_{3/2}(z_{1c})} \nonumber \\ &+\int_{T_c}^T\frac{1}{k_BT}\frac{d\Delta}{ dT}dT  
		 - \left[\frac{\mu-\varepsilon_{0}-\Delta}{k_BT} + \frac{\Delta_c}{k_BT_c} \right],
	\end{split}
	\label{Smn}
\end{equation}
where $S_b(T^-_c)/N_bk_B$ comes from (\ref{Smntmtc}) evaluated at $T_c$.


Again, we analyze the special cases: when
$\Delta = 0$ 
\begin{equation}
	\begin{split}
		&\frac{S_b(T)}{N_bk_B}= \frac{5}{2}\frac{g_{5/2}(z_1)}{g_{3/2}(z_1)} -\frac{5}{2}\frac{g_{5/2}(z_{1c})}{g_{3/2}(z_{1c})}   \\ 
		& + \frac{S_b(T_c)}{N_b k_B} - \frac{\mu-\varepsilon_{0}}{k_BT},
	\end{split}
	\label{Smn02}
\end{equation}
and when $\Delta  = \Delta_0$ constant
\begin{equation}
	\begin{split}
		&\frac{S_b(T)}{N_bk_B}= \frac{5}{2}\frac{g_{5/2}(z_1)}{g_{3/2}(z_1)} -\frac{5}{2}\frac{g_{5/2}(z_{1c})}{g_{3/2}(z_{1c})}   \\ 
		& + \frac{S_b(T_c)}{N_b k_B} - \left[\frac{\mu-\varepsilon_{0}-\Delta_0}{k_BT} + \frac{\Delta_0}{k_BT_c} \right].
	\end{split}
\label{SD_no0}
\end{equation}

\begin{figure}[htb]
\centerline{\epsfig{file=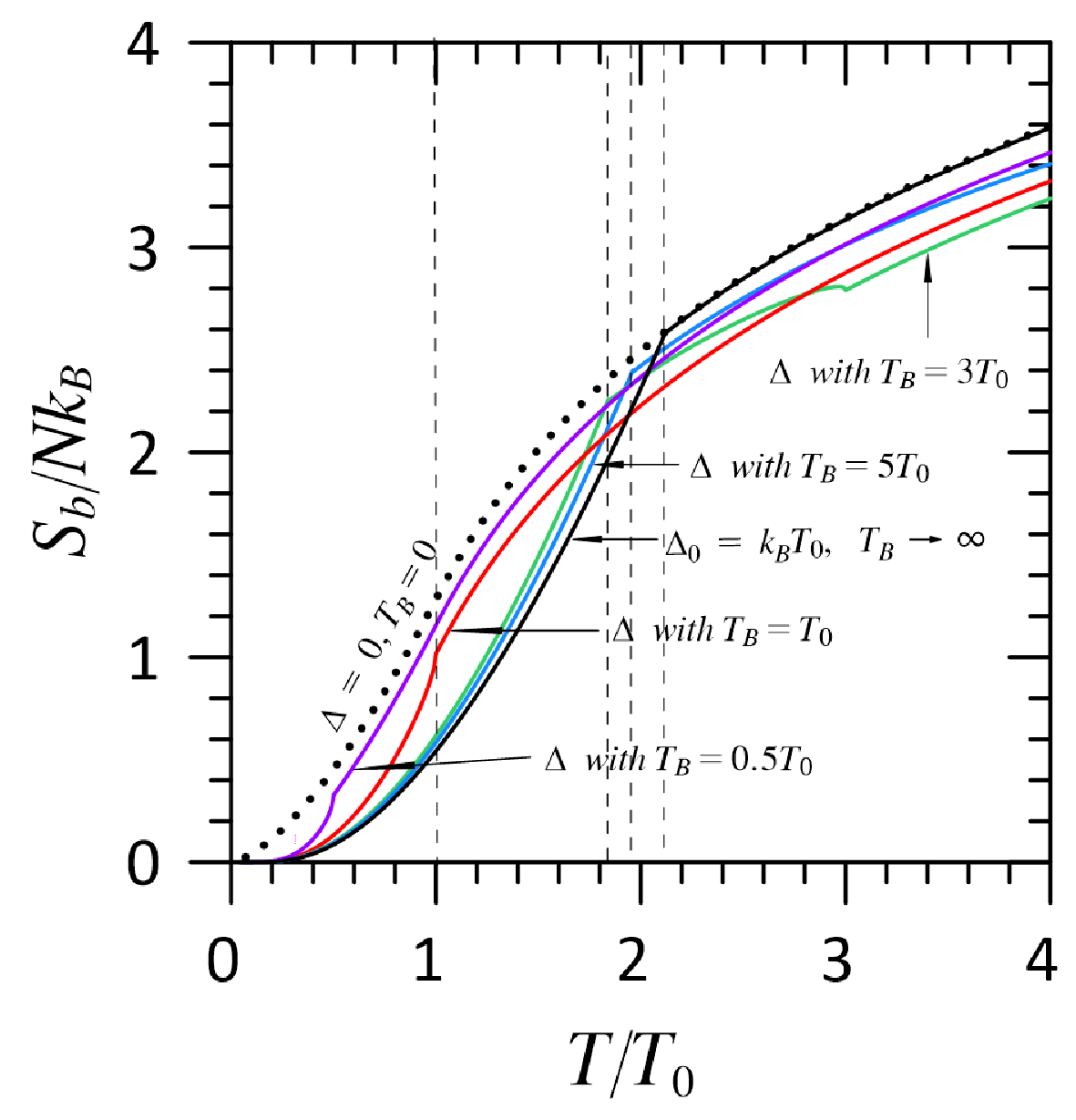,height=3.0in,width=3.5in}}
\hspace{-1.cm}
\caption{Normalized entropy. The dotted line represents the case without a gap, or $T_B=0$. The black line describes the entropy with a constant gap equal to $k_B\,T_0$, with $T_c = 2.12\,T_0$. For the case of a BCS gap with $T_B = 0.5\,T_0$, and $T_0$ we obtain that $T_c = T_0$; and when $T_B = 3\,T_0$ and $5\,T_0$ we get $T_c = 1.85\,T_0$ and $T_c = 1.97\,T_0$ respectively.}
\label{fig:S}
\end{figure}
Fig. \ref{fig:S} shows the behavior of the entropy as a function of temperature. The gap used is described by Eq.~(\ref{gapTB}), with $\Delta_0 = k_B\,T_0$. For every $T_B \leq T_0$ the critical temperature is $T_0$; when $T_B \rightarrow \infty $, that is, a constant gap $\Delta_0$, the critical temperature is $2.12\,T_0$. On the other hand, for $T_B > T_0$ the BEC critical temperature satisfies $T_0 < T_c < T_B$. For example, for $T_B=3\,T_0$ and $5\,T_0$ the BEC critical temperatures are $1.85\,T_0$ and $1.97\,T_0$ respectively. For every case, the entropy increases with temperature and has a kink at $T_B$. At $T_c$ the entropy has a change in concavity, but it is continuous, which reinforces the idea that the Bose-Einstein condensation has a continuous transition. We should also note that for $T_B \geq T_0$ a kink appears in $T_c$.

\subsection{Helmholtz free energy}

The Helmholtz free energy $F$ is given by 
\begin{equation}
	\frac{F}{Nk_B}=\frac{U}{Nk_B}-T\frac{S}{Nk_B}.
\end{equation}

\subsubsection{Boson Helmholtz free energy}

For a boson gas at $T\leq T_c$, substituting the form of the internal energy given in (\ref{UkBT}), with $\mu=\mathbf{\varepsilon}_{0}$, and after using the expression (\ref{Smntmtc}) for entropy we obtain
\begin{equation}
	\begin{split}
		\frac{F_b}{N_bk_BT_0}=\frac{\mathbf{\varepsilon }_{0}}{k_BT_0}-\left(\frac{T_0}{T_c} \right)^{3/2}
		\left[ \left(\frac{T}{T_0}\right)^{5/2}\frac{g_{5/2}(e^{-\beta\Delta})}{g_{3/2}(z_{1c})} \right. \\
		\left. +\frac{T}{T_0}\int_{0}^{T}\left(\frac{T}{T_0}\right)^{1/2}\frac{1}{k_BT_0}\frac{d\Delta}{dT}\frac{g_{3/2}(e^{-\beta\Delta})}{g_{3/2}(z_{1c})} dT \right] . \label{Fmntmenortc}
	\end{split}
\end{equation}

To find the relationship between $k_BT_0$ and Fermi's energy, we make use of Eq. (\ref{T0}), so 
\begin{equation}
	E_F=k_BT_F=\left(\frac{6\pi^2n_e}{g}\right)^{2/3}\frac{\hbar^2}{2m_e},
	\label{EFeq24Pathria}
\end{equation}
(Ref. \cite{Pathria} Section 8.1, Eq.(24)), where $n_e$ is the density of the electrons. Thus, since $n_e = 2 n_b$ and $m_b=2m_e$, we get
\begin{equation}
	T_0=0.218\,T_F.
	\label{T002TF}
\end{equation}
We can now express the boson Helmholtz free energy in units of the Fermi energy. For the case of the constant gap, using (\ref{eq:Tc/TF}) and (\ref{T002TF}) in (\ref{Fmntmenortc}) we get
	\begin{flalign}
		\frac{F_b}{N_bk_BT_F} &= 
	 -\frac{(T/T_F)^{5/2}}{0.102\,\zeta(3/2)}g_{5/2}(\exp[-\beta( \Delta_0)]) \notag \\
		&+ \frac{\varepsilon_0}{k_BT_F}.
	\end{flalign}


\subsubsection{Fermion Helmholtz free energy}

The Helmholtz free energy for the unpaired electron gas $F_e$  
(Ref.\cite{Pathria}, Section 8.1, Eq. (11)), in units of $N_ek_BT_F$, is given by
\begin{equation}
\frac{F_e}{N_ek_BT_F}=\frac{T}{T_F}\left[\beta\mu_e-\frac{f_{5/2}(\exp[\beta\mu_e])}{f_{3/2}(\exp[\beta\mu_e])}\right]. \label{FeskbTF}
\end{equation}
From \cite{Pathria}, we can use
\begin{equation}
	\frac{N_e}{V}=\frac{(2\pi m_ek_BT)^{3/2}g}{h^3}f_{3/2}(\exp[\beta\mu_e])
	\label{mueeq4Pathria}
\end{equation}
to obtain $\mu_e$, where $g$ is the spin degeneration. 
Then, using Eq. (\ref{EFeq24Pathria}) we can rewrite the last equation as 
\begin{equation}
	\frac{4}{3\pi^{1/2}}=\left(\frac{T}{T_F}\right)^{3/2}f_{3/2}(\exp[\beta\mu_e]),
\end{equation}
from where the chemical potential for the fermion gas can be extracted numerically. Notice that the last expression is independent of $g$.


\section{Condensation energy}

\begin{figure}[htb]
	\vspace{-0.0cm}
		\hspace{-0.5cm}
	\centerline{\epsfig{file=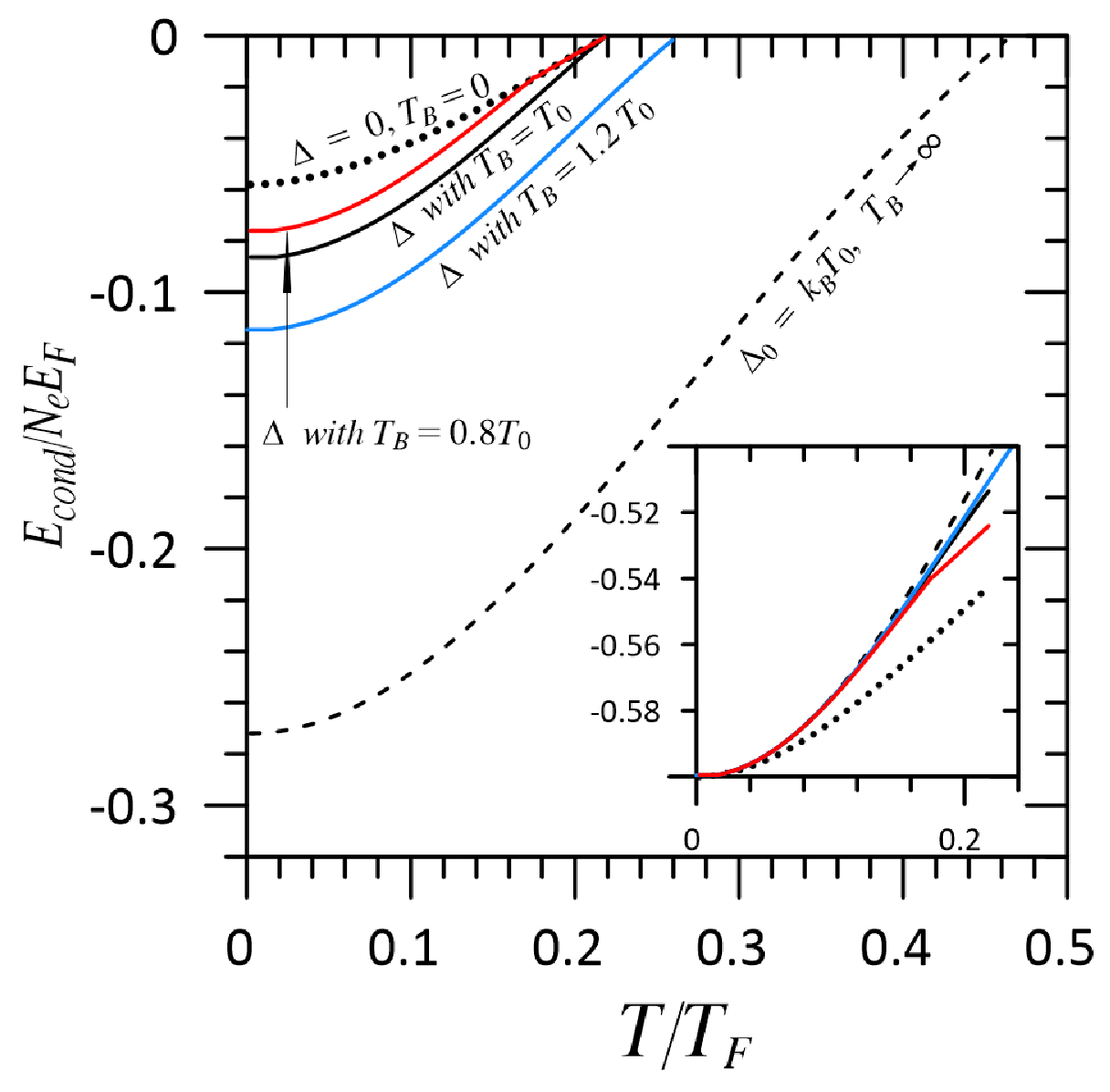,height=3.0in,width=3.5in}}
	\vspace{0.0cm}
	\caption{Condensation energy calculated using the Boson-Fermion formalism, Eq. (\ref{Econd F-FeTc}), referred to its value at $T_c$ in units of $E_F$. The gap used is the BCS gap, Eq. (\ref{gapTB}), with $\Delta_0=k_BT_0$ and several values of $T_B$: $T_B=0$ ($\Delta=0$) and $T_c =0.218\,T_F$ (dotted line); $T_B \rightarrow \infty $ ($\Delta=\Delta_0$) and $T_c =0.463\,T_F$ (dashed line); $T_B=0.8\,T_0$ (full red line); $T_B=T_0$ (full black line), both with $T_c =0.218\,T_F$ and $T_B= 1.2\,T_0$ (full blue line) with $T_c =0.261\,T_F$. In every case $\varepsilon_0=0$. The inset shows the condensation energy for the same cases without referring to its value at $T_c$.}
	\label{fig:EcondFb-Fe}
\end{figure}

The condensation energy is defined as the difference between the Helmholtz free energies of the superconducting and normal states, that is, $E_{cond}(T) = F_s - F_n$. \\


\subsection{Boson-Fermion condensation energy}

In the Boson-Fermion formalism, the normal state is a gas of $N$ interacting electrons, where only those that are   inside the Debye shell around the Fermi surface have the possibility of forming Cooper pairs,
so the number of paired electrons is a small fraction $f N_{\omega}$ of the number of electrons $N_{\omega}$ inside the Debye shell, with $f<1$ determined by making $T_c$ equal to the superconducting experimental values. 
On the other hand, the free energy $F_s$ of the superconducting state comes from the sum of the free energies of $f N_{\omega}/2$ Cooper pairs considered as composite bosons, which we denote as  $F_{f N_{\omega}/2}$, plus that of the unpaired electrons $F_{(1-f)N_{\omega}}$ within the Debye shell, in addition to the free energy of the electrons outside of it.  Writing $F_s= F_{f N_{\omega}/2} + F_{(1-f)N_{\omega}} + F_{no}$, where $F_n = F_{ni} + F_{no}$ with $F_{ni}$ the HFE of the electrons inside and  $F_{no}$ of those outside the Debye shell, respectively, we have that the condensation energy using the Boson-Fermion formalism is
	\begin{eqnarray}  
	E_{cond}(T) &=& F_s - F_n  \nonumber \\
	 &=&  ( F_{f N_{\omega}/2} + F_{(1-f)N_{\omega}} + F_{no})-(F_{ni} + F_{no}) \nonumber \\ 
 &=& (F_{(1-f)N_{\omega}}-F_{ni}) + F_{f N_{\omega}/2} \nonumber\\ 
	&=&  F_{f N_{\omega}/2}-F_{f N_{\omega}}.
	\label{Fs menos Fn}
\end{eqnarray} 
Taking as reference the condensation energy value at $T_c$, we have
\begin{eqnarray}
	E_{cond}= F_s(\Delta)-F_n&-&\left[F_{sc}(\Delta_c)-F_{nc}\right] \nonumber\\
	=F_{fN_\omega/2}(\Delta)-F_{fN_\omega}&-&\left[F_{fN_{\omega/2}c}(\Delta_c)-F_{fN_{\omega}c}\right].
\label{Econdfc}
\end{eqnarray}
\noindent
where the subscript $c$ tells us that the parameter is evaluated at $T_c$. 
Defining $b\equiv fN_\omega/2$ in Eq. (\ref{Fmntmenortc}) and $e\equiv fN_\omega$ in Eq. (\ref{FeskbTF}) we can rewrite (\ref{Econdfc}) as 
\begin{eqnarray}
	E_{cond}= F_b(\Delta)-F_e-\left[F_{bc}(\Delta_c)-F_{ec}\right].
	\label{Econd F-FeTc}
\end{eqnarray}

Figure \ref{fig:EcondFb-Fe} shows the condensation energy (\ref{Econd F-FeTc}) for $\varepsilon_0=0$ in units of $E_F$. The energy gap used is described in (\ref{gapTB}) with $\Delta_0=k_BT_0$ and several values of $T_B$, which are $T_B=0$, i.e., $\Delta=0$, where $T_c =0.218\,T_F$; $T_B \rightarrow \infty $, i.e., $\Delta=\Delta_0$ where $T_c =0.463\,T_F$; $T_B=0.8\,T_0$ and $T_B=T_0$, where $T_c =0.218\,T_F$ and $T_B=1.2\,T_0$, where $T_c =0.261\,T_F$. 

The inset in Fig. \ref{fig:EcondFb-Fe} shows $E_{cond}$ for the cases mentioned, but using Eq. (\ref{Fs menos Fn}), which is the condensation energy without taking its value in $T_c$ as a reference. Note that in every case the condensation energy has the same value for $T=0$, which corresponds to the internal energy of the unpaired electrons $U_e(T=0)= -3/5 N_eE_F$.


\subsection{Ginzburg-Landau-BCS condensation energy}
In the Ginzburg-Landau Theory, the condensation energy is given by \cite{GL}
\begin{equation}
E_{GL} = F_s-F_n = -V \left(\frac{\alpha^2}{2b}\right)T_c^2\left(1-\frac{T}{T_c}\right)^2,
\label{EconGL}
\end{equation}
where $F_s$ and $F_n$ are the Helmholtz free energies in the superconducting and normal states, $\alpha$ and $b$ are phenomenological parameters, and $V$ is the volume.

For $T=0$ the BCS theory gives
\begin{eqnarray}
E_{BCS} &=& F_s-F_n =-\frac{D(E_F)}{2}\Delta^2(0) \nonumber \\
&=&-\frac{D(E_F)}{2}\left(1.764\,k_BT_c\right)^2,
\end{eqnarray}
 where $\Delta(0)$ is the BCS gap for $T=0$ and $D(E_F)$ is the density of states at the Fermi energy level (Ref. \cite{Tinkham}, pags. 58 and 63) given by
 \begin{equation}
 	D(E_F)= \ \frac{gV}{4\pi^{2}}\left(\frac{2m_e}{\hbar^{2}}\right)^{3/2}E_F^{1/2}
 	\label{DEFg}
 \end{equation}
using $g=1$ (for one spin).

Combining the last two equations, and taking $V \left(\alpha^2/2b\right)=\left(D(E_F)/2\right)\left(1.764\,k_B\right)^2$ we get the 
``Ginzburg-Landau-BCS condensation energy"
\begin{equation}
E_{GL-BCS}	= -\frac{D(E_F)}{2}\left(1.764\,k_BT_c\right)^2\left(1-\frac{T}{T_c}\right)^2. \label{GLBCS}
\end{equation}
By using the Ginzburg-Landau theory, we are able to obtain an expression valid from $T=0$, i. e., BCS, to $T=T_c$.


\section{Universal behavior of $E_{cond}/\gamma_0$}
\begin{figure}[htb]
	\vspace{0.20cm}
	\centerline{\epsfig{file=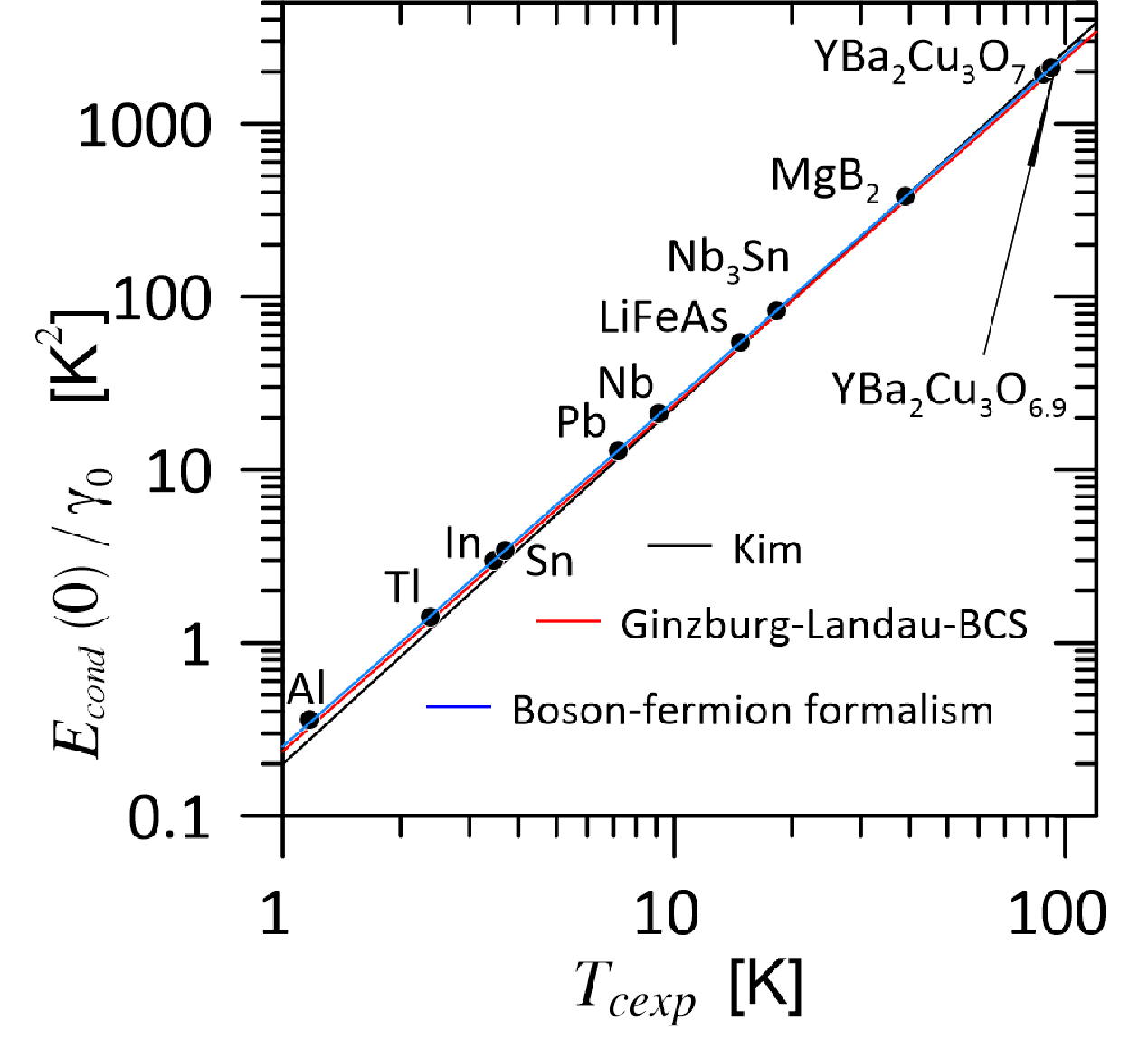,height=3.0in,width=3.50in}}
	\vspace{.2cm}
\caption{
	The absolute value of the condensation energy at $T=0$ divided by the Sommerfeld constant as a function of the critical temperature of the Bose-Einstein condensation of various superconducting materials using the Boson-Fermion formalism, Eqs. (\ref{Econd F-FeTc}) and  (\ref{gamma0}) (black dots and blue line), and the Ginzburg-Landau-BCS expression, Eqs.  (\ref{G-L-BCSgamma}) and (\ref{gamma0}) (red line). The result obtained by J. S. Kim (Ref. \cite{Kim-Tam}, Fig. 5) is also included (black line). }
	\label{fig:EcondgammaVsTcb}
\end{figure}

{
	
	\begin{table*}[ht]
			\caption{CONDENSATION ENERGY 
			}
			\setlength{\tabcolsep}{4pt}
			\begin{tabular}{|c|c|c|c|c|c|c|c|c|c|c|c|} \hline \hline
				& \multicolumn{3}{c|}{$E_{cond}(0)$ [mJ/mol]}
				& \multicolumn{3}{c|}{$\gamma_0$ [mJ/mol K$^2$] }
				& \multicolumn{3}{c|}{$E_{cond}(0)/\gamma_0$ [K$^2$/mol]} 
				& Tc [K]
				&f \\ 
				& Exp  
				& B-F$^{\diamond}$
				& GL-BCS$^{\dagger}$
				& Exp.  
				& B-F$^{\ddagger}$
				& BCS$^{\ast}$
				& Exp.  
				& B-F$^{\ast}$$^{\ast}$ 
				& GL-BCS$^{\circ}$
				& Exp.
				&[10$^{-5}$]\\ \hline
				Al  & -0.438& -0.444168 & -0.295249 & 1.36 & 1.2623 & 0.912132 & -0.322&-0.3519
				& -0.324 &1.17& 1.9354 \\ 
				Tl  & -2.12   & -2.59544& -1.77375 & 1.47 & 1.81562 & 1.31322  &-1.442&-1.42951
				& -1.35069 &2.39& 21.2454\\
				In  & -5.00   &-5.0995  &-3.48386 & 1.66 & 1.70588 & 1.23069 & -3.012 & -2.98936 & -2.83081 &3.46&59.4644\\
				Sn  & -6.26   & -7.02264  & -5.72385 & 1.76 & 2.04697 & 1.75866 & -3.557 & -3.43074 & -3.25467
				&3.71& 34.5289 \\ 
				Pb  & -46.8   & -24.3111  & -18.3477 & 2.99 & 1.88582 & 1.49844 & -15.65 & -12.8915 & -12.2445
				&7.196& 114.62\\ 
				Nb  & -164.6  & -27.8644 &-13.2574 & 7.80 & 1.33204 & 0.666746 & -21.10 & -20.9185 & -19.88
				&9.17 &18.6497 \\ \hline
				LiFeAs (111) & -423 & -435.192  & -52.6928 & 9.67 & 5.37436 & 0.687778 & -43.74 & -80.9757 & -76.61
				&18 \cite{Nag}&9.51119\\
				Nb$_3$Sn  & -4757   &  -284.909 & -58.6483 & 53.4 & 3.42 & 0.740619 & -89.08 & -83.301 & -79.19
				&18.3 \cite{Guritanu}&273.206 \\
				MgB$_2$  & -738    & -288.884  & -17.4298 & 2.69 & 0.76 & 0.048462 & -274.3& -379.7 & -359.7
				&39 \cite{Budko2015}&47.3492 \\			
				YBa$_2$Cu$_3$O$_{7}$  & -60471 & -11583.9 & -2711.62 & 21.0 & 6.04 & 1.48758 & -2880& -1919 & -1823
				&87.8 \cite{YuxingWang} & 612.948 \\ 
				YBa$_2$Cu$_3$O$_{6.9}$  & -   &  -12769.8  & -2667.82 & - & 6.04 & 1.33298 & - & -2115 & -2001
				&92 \cite{Liang}&61.7404 \\
				\hline 
			\end{tabular} \label{tab:EcG}
			The experimental values 
			that do not have a citation 
			are taken from Table I of \cite{Kim-Tam}. $^{\diamond}$Boson-Fermion values calculated using Eq. (\ref{Econd F-FeTc}). $^{\dagger}$GL-BCS values calculated using Eq. (\ref{GLBCS}) . $^{\ddagger}$Using Eq. (\ref{gamma0}) for $g=2$. ${^\ast}$Using Eq. (\ref{gamma0}) for $g=1$. $^{\ast}$$^{\ast}$ Using Eq. (\ref{Econd F-FeTc}) and Eq. (\ref{gamma0}) for $g=2$. $^{\circ}$Using  Eq. (\ref{G-L-BCSgamma}) and Eq. (\ref{gamma0}) for $g=1$.
		\end{table*}
	}

In this section we analyze the ratio between the condensation energy, obtained using the two methods described above, and 
the Sommerfeld's constant 
$\gamma_0$ for a wide range of superconductors, where $\gamma_0$ is the coefficient between the electronic specific heat and the temperature for $T\rightarrow 0$ 
usually obtained from experimental data, as in \cite{Kim-Tam}. However, in our calculations, we use an analytical expression for obtaining $\gamma_0$.

The specific heat at constant volume for low temperatures for a $3D$ free electron gas confined in a box is (Ref. \cite{Cetina}, Eq. (16))
\begin{equation}
	C_{\tiny V} = \frac{2}{3}\pi^{2}D(E_F)k_B^{2}T= \gamma_0T,
\end{equation}
from where
\begin{equation}
	\gamma_0=\frac{2}{3}\pi^2D(E_F)k_B^2. \label{gamma0}
\end{equation}

Dividing the Ginzburg-Landau-BCS condensation energy, Eq. (\ref{GLBCS}), by Eq. (\ref{gamma0}) for the case $g=1$, we get
\begin{equation}
\frac{E_{GL-BCS}}{\gamma_0}=	\frac{F_s-F_n}{\gamma_0} = -0.236\,T_c^2\left(1-\frac{T}{T_c}\right)^2
	\label{G-L-BCSgamma}.
\end{equation}

Figure \ref{fig:EcondgammaVsTcb} shows the absolute values of the condensation energy divided by $\gamma_0$ at $T=0$ as a function of the experimental values of $T_c$ for various conventional and  unconventional superconductors: first using the Boson-Fermion formalism,  Eq. (\ref{Econd F-FeTc}), (black dots), whose fit gives $E_{cond}/\gamma_0=0.252\,T_c^{1.997}$ (blue line) which is comparable to the fit $E_{cond}/\gamma_0=0.2\,T_c^{2.06}$ compiled  by J. S. Kim  (Ref. \cite{Kim-Tam}, Fig. 5) (black line) for the experimental values of conventional and unconventional superconductors.
Also, we plot the curve obtained using the Ginsburg-Landau-BCS expression Eq. (\ref{G-L-BCSgamma}) (red line curve), which gives $E_{GL-BCS}/\gamma_0 =0.236\,T_c^{2}$.
%
The reason why the three lines fall very close  
could be due to when we take the ratio of the condensation energy at $T=0$ to the Sommerfeld constant, the dependence on the density of states at the Fermi level, which is an intrinsic characteristic of each superconductor, is cancelled.
It is well known that the BCS theory is able to describe very well, at least, the conventional superconductors; hence, the dependence on the density of states vanishes when the ratio between the condensation energy at $T=0$ and the Sommerfeld constant is taken. On the other hand, the Boson-Fermion formalism has shown to be useful in describing the properties of conventional and some non-conventional superconductors, so when we take the same ratio using the corresponding condensation energy, Eq. (\ref{Fs menos Fn}), the dependence on the density of states is also canceled.





On the other hand, we are interested in obtaining the number $f N_{\omega}$ of the paired electrons inside the Debye shell, 
where $N_{\omega}$ is given by
\begin{equation}
	N_{\omega}=\frac{V}{4\pi^{2}}\left(\frac{2m_e}{\hbar ^{2}}\right)^{3/2}\int_{E_F -\hbar \omega_D}^{E_F +\hbar \omega_D} \frac{E^{1/2}dE}{e^{\beta (E-\mu)}+1},
\end{equation}
with $\omega_D$ the Debye frequency (Ref. \cite{Tinkham}, pag. 45). For $T=0$
\begin{equation}
	N_{\omega}=\frac{V}{4\pi^{2}}\left(\frac{2m_e}{\hbar ^{2}}\right)^{3/2}\int_{E_F -\hbar \omega_D}^{E_F}E^{1/2}dE.
\end{equation}
After integrating and taking $n_{\omega} = N_{\omega}/V$,
\begin{equation}
	n_{\omega}=\frac{1}{6\pi^{2}}\left(\frac{2m_e}{\hbar ^{2}}\right)^{3/2}E_F^{3/2}\left[1-\left(1 - \frac{\hbar\omega_D}{E_F}\right)^{3/2}\right].
	\label{nw}
\end{equation}



Then, the values of $f$ are obtained using $n_b = fn_\omega/2$ in Eq. (\ref{Tc}) and solving for $\beta_c$, so
\begin{equation}
	\begin{split}	
		T_{c}=
		0.218033 \ &(4.23124\times10^{-15}
		(fn_{\omega})^{2/3} \mbox{m}^2\mbox{K} ) \\
		& \times \left( \frac{\zeta (3/2)}{g_{3/2}(z_{1c})}\right)^{2/3},
		\label{Tcbf}
	\end{split}
\end{equation}
where the experimental value of $T_c$ is taken as the critical temperature of a system of $fn_\omega/2$ bosons. 



In Table \ref{tab:EcG} we present 
the condensation energy values calculated using both the Boson-Fermion formalism Eq. (\ref{Econd F-FeTc}) and the GL-BCS theory Eq. (\ref{GLBCS}), alongside the experimental values for various superconducting materials. We also show the calculated ratios of the previous quantities and the Sommerfeld constant Eq. (\ref{gamma0}), and the calculated values of $f$ using Eq. (\ref{Tcbf}).
It is important to note that the calculated values are very close to the experimental values for conventional materials, which is not the case for high-temperature materials; nevertheless, the calculated and experimental $E_{cond}/\gamma_0$ ratios are very similar (see Table \ref{tab:EcG}).


\section{Conclusions}



Modeling superconductors as a mixture of Cooper pairs (composite bosons) plus unpaired electrons, we give their condensation energy as the difference between the Helmholtz free energies of the superconducting and the normal state. 
	
In the Boson-Fermion formalism, the normal state 
is an original attractive electron gas, while the superconducting state is formed by the condensed Cooper pairs taken as composite bosons, coming from
a fraction $f < 1$ of electrons inside the Debye shell, plus those electrons inside and outside the Debye shell that are not able to pair. 
The condensation energy is the difference between the Helmholtz free energies of paired electrons taken as bosons and that of the same electrons but unpaired taken as fermions, both sets within the Debye shell.
For many superconductors ranging from conventional to high critical temperature, using the Boson-Fermion formalism, we show that the ratio of the condensation energy at $T=0$ divided by the corresponding Sommerfeld constant $E_{cond}(T=0)/\gamma_0$, as a function of the superconducting critical temperature taken as the BEC critical temperature, satisfies a universal behavior that goes like $0.252\,T_c^{1.997 }$, 
while 
from the Ginsburg-Landau-BCS expression 
goes like $0.236\,T_c^2$.  With both methods
the  $E_{cond}(T=0)/\gamma_0$ ratio are in 
agreement with the experimental fit reported by Kim \cite{Kim-Tam} for several conventional and non-conventional superconductors, where the ratio is 0.2\,$T_c^{2.06}$, as well as with the experimental results recently reported by Tallon \cite{Tallon2026} for overdoped cuprate superconductors.
This universal behavior of
$E_{cond}/\gamma_0$ at $T=0$ 
for a wide range of superconducting materials could be due to the fact that
its dependence on the density of states, an innate characteristic of every material, disappears when the ratio is taken.



Finally, we notice that the condensation energy referenced to its value at $T_c$ has the highest absolute values, for $T < T_c$, for the case of a constant gap and the minimum ones for the case without a gap, which is the same behavior observed for the values obtained for $T_c$.

%
	
%
	

\textbf{Acknowledgments}
IC thanks SECIHTI (Mexico) for Postdoc grant EPA1 \# 869450. PS and MAS
thank PAPIIT-DGAPA-UNAM for grant IN114523.

\end{document}